# Surface Impedance Determination via Numerical Resolution of the Inverse Helmholtz Problem

Danish Patel[*], Prateek Gupta[*], and Carlo Scalo[†]

*Purdue University, West Lafayette, Indiana 47907, USA*

**Assigning boundary conditions, such as acoustic impedance, to the frequency domain thermoviscous wave equations (TWE), derived from the linearized Navier-Stokes equations (LNSE) poses a Helmholtz problem, solution to which yields a discrete set of complex eigenfunctions and eigenvalue pairs. The proposed method – the inverse Helmholtz solver (iHS) – reverses such procedure by returning the value of acoustic impedance at one or more unknown impedance boundaries (IBs) of a given domain, via spatial integration of the TWE for a given real-valued frequency with assigned conditions on other boundaries. The iHS procedure is applied to a second-order spatial discretization of the TWEs on an unstructured staggered grid arrangement. Only the momentum equation is extended to the center of each IB face where pressure and velocity components are co-located and treated as unknowns. The iHS is finally closed via assignment of the surface gradient of pressure phase over the IBs, corresponding to assigning the shape of the acoustic waveform at the IB. The iHS procedure can be carried out independently for different frequencies, making it embarrassingly parallel, and able to return the complete broadband complex impedance distribution at the IBs in any desired frequency range to arbitrary numerical precision. The iHS approach is first validated against Rott's theory for viscous rectangular and circular ducts. The impedance of a toy porous cavity with a complex geometry is then reconstructed and validated with companion fully compressible unstructured Navier-Stokes simulations resolving the cavity geometry. Verification against one-dimensional impedance test tube calculations based on time-domain impedance boundary conditions (TDIBC) is also carried out. Finally, results from a preliminary analysis of a thermoacoustically unstable cavity are presented.**

## I. Background

Acoustic impedance in a compressible fluid is defined as the ratio of the Fourier-transformed pressure to velocity evaluated at a given point in space, $\mathbf{x} = \{x_1, x_2, x_3\}$, and angular frequency, $\omega$. Hereafter, the following convention is adopted for pressure and velocity fluctuations, respectively

$$p'(\mathbf{x}, t) = \Re\left[\hat{p}(\mathbf{x};\omega)\, e^{j\omega t}\right], \quad \mathbf{u}'(\mathbf{x}, t) = \Re\left[\hat{\mathbf{u}}(\mathbf{x};\omega)\, e^{j\omega t}\right], \tag{1}$$

where the superscript $(')$ indicates fluctuations about a base state and $(\hat{\,})$ the corresponding complex amplitudes. The same convention is applied to the rest of the fluctuating quantities as discussed in §III. An acoustic impedance can be associated to each velocity component, $\hat{\mathbf{u}}(\mathbf{x};\omega) = \{\hat{u}_1(\mathbf{x};\omega), \hat{u}_2(\mathbf{x};\omega), \hat{u}_3(\mathbf{x};\omega)\}$, hence yielding the definition

$$Z_i(\mathbf{x};\omega) \equiv \frac{\hat{p}(\mathbf{x};\omega)}{\hat{u}_i(\mathbf{x};\omega)}. \tag{2}$$

The acoustic admittance is defined as the reciprocal of the impedance, $Y_i(\mathbf{x};\omega) = Z_i(\mathbf{x};\omega)^{-1}$, and can also be defined in vector notation, $\mathbf{Y}(\mathbf{x};\omega) \equiv \hat{\mathbf{u}}(\mathbf{x};\omega)/\hat{p}(\mathbf{x};\omega)$. Usually, impedance is defined based on the velocity component taken along the normal $\hat{\mathbf{n}}$ to a given surface $\mathcal{S}$,

$$Z_n(\mathbf{x};\omega) = \frac{\hat{p}(\mathbf{x};\omega)}{\hat{u}_n(\mathbf{x};\omega)}, \qquad \forall\ \mathbf{x} \in \mathcal{S} \tag{3}$$

---

[*]Graduate Research Assistant, School of Mechanical Engineering
[†]Assistant Professor, School of Mechanical Engineering, scalo@purdue.edu

where $u_n(\mathbf{x};\omega) = \hat{\mathbf{u}}(\mathbf{x};\omega) \cdot \hat{\mathbf{n}}(\mathbf{x})$, and will be hereafter referred to as normal impedance, related to normal admittance via $Y_n = Z_n^{-1}$. The impedance is a complex quantity, whose real and imaginary parts are referred to as resistance, $R(\mathbf{x};\omega) = \Re[Z(\mathbf{x};\omega)]$, and reactance, $X(\mathbf{x};\omega) = \Im[Z(\mathbf{x};\omega)]$, respectively. The base impedance defined as $Z_0 = \rho_0 \, a_0$, where $\rho_0$ and $a_0$ are the base density and speed of sound, is often used to obtain the specific (or normalized) value of acoustic impedance and admittance, hereafter respectively indicated as $Z_* = Z/\rho_0 \, a_0$ and $Y_* = \rho_0 \, a_0 \, Y$.

The normal resistance, $R_n$, explicitly appears in the isentropic expression of the cycle-averaged acoustic power, $d\dot{W}_{ac}$, transmitted through an infinitesimal area, $dA$, centered about a location $\mathbf{x}$ on the surface $\mathcal{S}$ of assigned impedance,

$$d\dot{W}_{ac}/dA = \frac{\omega}{2\pi} \int_0^{\frac{2\pi}{\omega}} p'(\mathbf{x};\tau) u'_n(\mathbf{x};\tau) d\tau = \frac{1}{2} \Re\left[\hat{p}^*(\mathbf{x};\omega)\hat{u}_n(\mathbf{x};\omega)\right] = \frac{1}{2} R_n(\mathbf{x};\omega) \hat{u}_n(\mathbf{x};\omega) \, \hat{u}_n^*(\mathbf{x};\omega) \quad (4)$$

valid for a time-harmonic fluctuation of type (1). For a broadband periodic fluctuating field, the overall acoustic power can be obtained as the summation of tonal contributions (4) invoking orthogonality among different harmonics.

The result in (4) seems to suggest that the imaginary part of the impedance does not affect the acoustic power. This can be easily confuted with the following thought experiment. Let $\mathcal{S}$ be a planar surface of assigned normal resistance $\Re[Z_n] = \rho_0 a_0$. Assigning a zero reactance, $\Im[Z_n] = 0$, allows a plane wave of (any) frequency $\overline{\omega}$ traveling along the direction of the normal $\hat{\mathbf{n}}$ to seamlessly cross the surface $\mathcal{S}$. On the other hand, assigning a reactance $\rho_0 \, a_0/\Im[X_n] = 0$ for $\omega \to \overline{\omega}$ to the same surface entails a complete reflection of the same wave. In the latter case, the surface acts as a purely reflective boundary at that frequency. In the former case, the acoustic power transmitted through an infinitesimal surface element is $d\dot{W}_{ac} = \frac{1}{2}\rho_0 \, a_0 \hat{u}_n(\mathbf{x};\omega) \, \hat{u}_n^*(\mathbf{x};\omega) dA$, in the latter case it is zero.

Assigning the spatial distribution of the broadband acoustic impedance (2) at a given surface retains all the necessary information to characterize the two-way acoustic coupling between waves on either side of the surface, provided wave propagation can be locally approximated as isentropic, that is, with one degree of thermodynamic freedom. Under non-isentropic conditions, more information needs to be specified to account for the second degree of thermodynamic freedom, which could be provided, for example, in the form of the ratio between the complex amplitudes of pressure and temperature or density, the latter not necessarily equal to $a_0^2$ under such conditions. We note that no isentropic-flow approximations are made in the iHS formulation (or when adopting the definition of acoustic impedance (2) itself) and the aforementioned ratios are readily retrievable as an output of the solver.

## II. Motivation and Previous Work

The acoustic properties of a system are typically investigated via Helmholtz solvers, which rely on an eigenvalue formulation (8), hence yielding a discrete set of complex eigenvalues and eigenvectors containing frequency and growth rate, and waveform shape information. Such approach requires homogeneous boundary conditions to be assigned. In an inviscid formulation, only the normal impedance (3) needs to be specified at all boundaries to close the eigenvalue problem. When adopting a viscous formulation, additional boundary conditions, related to the two tangential components of velocity and temperature, need to be specified: they can be in the form of homogeneous Dirichlet or Neumann conditions such as slip/no-slip for the velocity and isothermal/adiabatic for the temperature fluctuations, or, for example, be expressed as ratios of the frequency-transformed pressure to the tangential components of velocity (tangential impedance) or temperature. A classic eigenvalue approach therefore cannot be used to *calculate* the spatial distribution of the broadband impedance, for example, at the orifice of an acoustically absorptive cavity or the inlet of a nozzle. Knowing such quantity and applying it as a boundary condition to a flow simulation carried out only on one side of the impedance surface (e.g. high-fidelity time-domain simulation of the flow grazing such cavity or in a combustion chamber upstream the nozzle), would spare the need to resolve the flow on the other side of the same surface (e.g. in the cavity or in the nozzle). To achieve this goal, a novel methodology we call the 'inverse Helmholtz solver' or iHS has been developed and will be outlined in the present manuscript.

The iHS methodology presented herein does *not* belong to the class of inverse problems in acoustics that are concerned with the reconstruction of one or multiple acoustic sources from far field acoustic measurements.[1,2] It is more appropriately classifiable as an inverse Cauchy problem [3] applied to the thermoviscous-wave equations (TWE), where one or multiple unknown boundary conditions are determined via the assign-ment of the value of the angular frequency $\omega$ (as in the convention (1)) and at least one other boundary condition. Such problems are typically solved using iterative techniques[4,5] and rely on an initial guess for the unknown boundary condition that is updated until a given condition is satisfied within a tolerance. Such methods are computationally intensive since they mandate the solution of a direct problem at each step, and often may not be able to satisfy the convergence condition to the desired accuracy.[6] The proposed method solves the inverse Cauchy problem applied to the compressible linearized Navier-Stokes equations via a one-time, direct resolution of a system of linear equations.

Inverse problems associated to acoustic source reconstruction or Helmholtz equation do not typically have a unique or well-defined solution, and may require the adoption of regularization techniques.[7,8] Grayzin et. al[4] applied Sommerfeld radiation conditions to the Helmholtz problem arriving at an over-constrained linear system of equations which was then solved using least squares. A meshless, direct method to determine unknown boundary conditions to the Helmholtz problem was proposed by Jin & Zheng[9] relying on the method of fundamental solutions and truncated singular value decomposition regularization. Piechowicz et. al[10–12] have developed a numerical scheme that relies on a boundary element method applied to the Helmholtz-Kirchoff equation to evaluate the acoustic impedance of obstacles and/or boundaries internal to closed domain from pressure information taken at a finite set of interior points. To the author's knowledge no prior work has tackled the evaluation of the broadband acoustic impedance (2) at one or multiple boundaries of a computational domain without the use of regularization techniques or iterative methods.

The iHS methodology is outlined in section §III of the present manuscript, where the linearized governing equations are first presented for a general isotropic compressible fluid (§III.A), yielding the set of thermoviscous wave equations (TWEs), which are recast in the form of an inverse Helmholtz problem with appropriate closure conditions (§III.C). The iHS is first validated against Rott's thermoviscous theory[13,14] in square and circular constant cross-sectional area ducts with both inviscid and viscous wave propagation assumption (section §IV). The impedance of a geometrically complex acoustically absorptive toy cavity is then considered (section §V), validated against time-domain pore-resolved unstructured Navier-Stokes simulations under both harmonic (§V.A) and pulsed (§V.B) excitation and with direct comparison against time-domain impedance boundary conditions (TDIBC) one-dimensional simulations .[15–17]

## III. Formulation of the inverse Helmholtz solver (iHS)

In this section, the inverse Helmholtz problem formulation (§III.C) is presented for the linearized compressible Navier-Stokes equations (LNSE) (§III.A) on an unstructured grid with a staggered variable arrangement discretized with second-order accuracy.

### III.A. Linearized Navier-Stokes Equations (LNSE)

Governing equations for a fully compressible flow are

$$\frac{\partial}{\partial t}(\rho) + \frac{\partial}{\partial x_m}(\rho u_m) = 0, \tag{5a}$$

$$\frac{\partial}{\partial t}(\rho e) + \frac{\partial}{\partial x_m}[u_m(\rho e + p)] = \frac{\partial}{\partial x_m}(u_i \tau_{im} - q_m), \tag{5b}$$

$$\frac{\partial}{\partial t}(\rho u_i) + \frac{\partial}{\partial x_m}(\rho u_i u_m) = -\frac{\partial}{\partial x_i}p + \frac{\partial}{\partial x_m}\tau_{im}, \tag{5c}$$

where $x_1$, $x_2$, and $x_3$ (or $x$, $y$, and $z$) are Cartesian coordinates, $u_i$ are the velocity field components in each of these directions, and $p$, $\rho$, and $e$ are, respectively, the pressure, density, and total energy per unit mass. The viscous shear stress, $\tau_{ik}$ and heat flux $q_m$ are defined as,

$$\tau_{im} = 2\mu \left[\frac{1}{2}\left(\frac{\partial u_m}{\partial x_i} + \frac{\partial u_i}{\partial x_m}\right)\right] + \lambda \frac{\partial u_k}{\partial x_k}\delta_{im}, \qquad q_m = -\kappa \frac{\partial T}{\partial x_m},$$

where $\mu$ and $\lambda$ are the first and second viscosity coefficients, respectively, and $\kappa$ is the thermal conductivity.

Decomposing all variables into a base state, denoted with the subscript '0' and a fluctuation, denoted with a prime, $\phi = \phi_0 + \phi'$, and substituting into (5) and linearizing, yields

$$\frac{\partial \rho'}{\partial t} + \frac{\partial(\rho_0 u'_k)}{\partial x_k} + \frac{\partial(u_{0,k}\rho')}{\partial x_k} = 0 \tag{6a}$$

$$\rho_0 \frac{\partial e'}{\partial t} + e_0 \frac{\partial \rho'}{\partial t} + \frac{\partial (\rho' e_0 u_{0,k})}{\partial x_k} + \frac{\partial (\rho_0 e' u_{0,k})}{\partial x_k} + \frac{\partial (\rho_0 e_0 u'_k)}{\partial x_k} + p_0 \frac{\partial u'_k}{\partial x_k}$$
$$+ p' \frac{\partial u_{0,k}}{\partial x_k} - \kappa \frac{\partial}{\partial x_k}\left(\frac{\partial}{\partial x_k} T'\right) = 2\lambda \frac{\partial u_{0,j}}{\partial x_j}\frac{\partial u'_k}{\partial x_k} + 2\mu \frac{\partial u_{0,k}}{\partial x_j}\frac{\partial u'_k}{\partial x_j} + 2\mu \frac{\partial u_{0,j}}{\partial x_k}\frac{\partial u'_k}{\partial x_j} \quad (6b)$$

$$\rho_0 \frac{\partial u'_i}{\partial t} + u_{0,i}\frac{\partial \rho'}{\partial t} + \frac{\partial (\rho_0 u'_i u_{0,k})}{\partial x_k} + \frac{\partial (\rho' u_{0,i} u_{0,k})}{\partial x_k} + \frac{\partial (\rho_0 u_{0,i} u'_k)}{\partial x_k} = -\frac{\partial p'}{\partial x_i} + \frac{\partial}{\partial x_k}\tau'_{ik}. \quad (6c)$$

For a perfect gas (calorically and thermally) with a homogeneous and quiescent base state, these equations simplify to the following conservation equations,

$$\frac{\partial p'}{\partial t} - \rho_0 R \frac{\partial T'}{\partial t} + p_0 \frac{\partial u'_k}{\partial x_k} = 0, \quad (7a)$$

$$\rho_0 \frac{\partial u'_i}{\partial t} + \frac{\partial p'}{\partial x_i} = \mu \frac{\partial}{\partial x_j}\left(\frac{\partial}{\partial x_j} u'_i\right) + (\mu + \lambda)\frac{\partial}{\partial x_i}\left(\frac{\partial u'_k}{\partial x_k}\right), \quad (7b)$$

$$\rho_0 c_v \frac{\partial T'}{\partial t} + p_0 \frac{\partial u'_k}{\partial x_k} - \kappa \frac{\partial}{\partial x_k}\left(\frac{\partial}{\partial x_k} T'\right) = 0. \quad (7c)$$

### III.B. Thermoviscous Wave Equations (TWE) and Numerical Discretization

Converting the equations (7) from time domain to frequency domain according to the convention (1), yields the thermoviscous wave equations, or TWE, not reported for brevity. TWE are discretized via a second-order spatial discretization. A staggered variable arrangement (figure 1) is used collocating thermodynamic field variables (pressure $\hat{p}$, specific energy $\hat{e}$, and hence temperature $\hat{T}$ and density $\hat{\rho}$) at the mesh element centroids and velocity field components $\hat{u}_i$ at the face centers.

Upon applying a spatial discretization and only homogeneous boundaries, the TWEs read

$$j\omega \underline{X}_0 + \underline{\underline{A}}_{NS} \underline{X}_0 = 0 \quad (8)$$

where $\underline{X}_0 = \left\{\hat{\underline{p}}, \hat{\underline{T}}, \hat{\underline{u}}_1, \hat{\underline{u}}_2, \hat{\underline{u}}_3\right\}^T$ and $\hat{\underline{\phi}}$ represents a column array collecting all discretized values of the complex amplitude of the generic thermo-fluid-dynamic quantity $\hat{\phi}$. Equation (8) describes a conventional eigenvalue problem yielding a discrete spectrum of frequencies (eigenvalues) and wave forms (eigenvectors) of the problem.

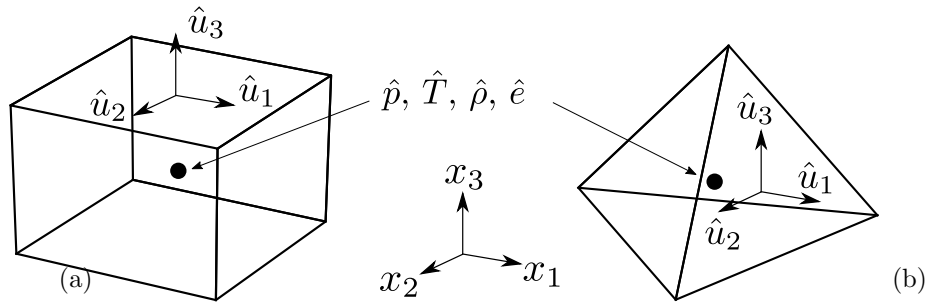

Figure 1: Staggered variable arrangement over hexahedral (a) and tetrahedral (b) cell. Thermodynamic quantities, $p$, $T$, $\rho$, $e$ and $c_v$, are stored at the cell centers (•), while velocity components, $u_i$, are located on face centers.

### III.C. Inverse Helmholtz Problem Formulation

To derive the inverse Helmholtz solver (iHS) formulation the TWE are first considered at a fixed or assigned real-valued frequency $\omega$. The iHS system reads:

$$j\omega \begin{pmatrix} \hat{\underline{p}} \\ \hat{\underline{T}} \\ \hat{\underline{u}}_1 \\ \hat{\underline{u}}_2 \\ \hat{\underline{u}}_3 \\ \hat{\underline{u}}_{1,\text{IB}} \\ \hat{\underline{u}}_{2,\text{IB}} \\ \hat{\underline{u}}_{3,\text{IB}} \\ \hat{\underline{p}}_{\text{IB}} \end{pmatrix} + \begin{pmatrix} \underline{\underline{A}}_{NS} & \underline{\underline{A}}_{\text{IB}} \\ \underline{\underline{\mathcal{M}}}_{\text{IB}} \\ \underline{\underline{\Phi}}_{\text{IB}} \end{pmatrix} \begin{pmatrix} \hat{\underline{p}} \\ \hat{\underline{T}} \\ \hat{\underline{u}}_1 \\ \hat{\underline{u}}_2 \\ \hat{\underline{u}}_3 \\ \hat{\underline{u}}_{1,\text{IB}} \\ \hat{\underline{u}}_{2,\text{IB}} \\ \hat{\underline{u}}_{3,\text{IB}} \\ \hat{\underline{p}}_{\text{IB}} \end{pmatrix} = \begin{pmatrix} \underline{b}_{\text{p}} \\ \underline{b}_{\text{T}} \\ \underline{b}_{\text{u}_1} \\ \underline{b}_{\text{u}_2} \\ \underline{b}_{\text{u}_3} \\ \underline{b}_{\text{u}_{1,\text{IB}}} \\ \underline{b}_{\text{u}_{2,\text{IB}}} \\ \underline{b}_{\text{u}_{3,\text{IB}}} \\ \underline{b}_{\text{IB}} \end{pmatrix} \quad (9)$$

where $\underline{\underline{A}}_{NS}$ represents the TWEs discretized in the interior of the domain and $\underline{\underline{\mathcal{M}}}_{\text{IB}}$ the frequency-transformed momentum equation extended to the center of each impedance boundary (IB) face where pressure and velocity components are collocated and treated as unknowns (figure 2b). The latter contribute to the TWE in the interior of the domain via the block $\underline{\underline{A}}_{\text{IB}}$. Having introduced four new unknowns at the IB $\{\hat{\underline{u}}_1, \hat{\underline{u}}_2, \hat{\underline{u}}_3\}^T, \hat{p}_{\text{IB}}$ but only three new equations, warrants the specification of a new set of linearly independent equations equal to the number of discrete faces at the IB, $n_{\text{IB}}$, to close the iHS problem. Such closure conditions are represented by the block $\Phi_{\text{IB}}$. Various closure strategies can be adopted and will be discussed.

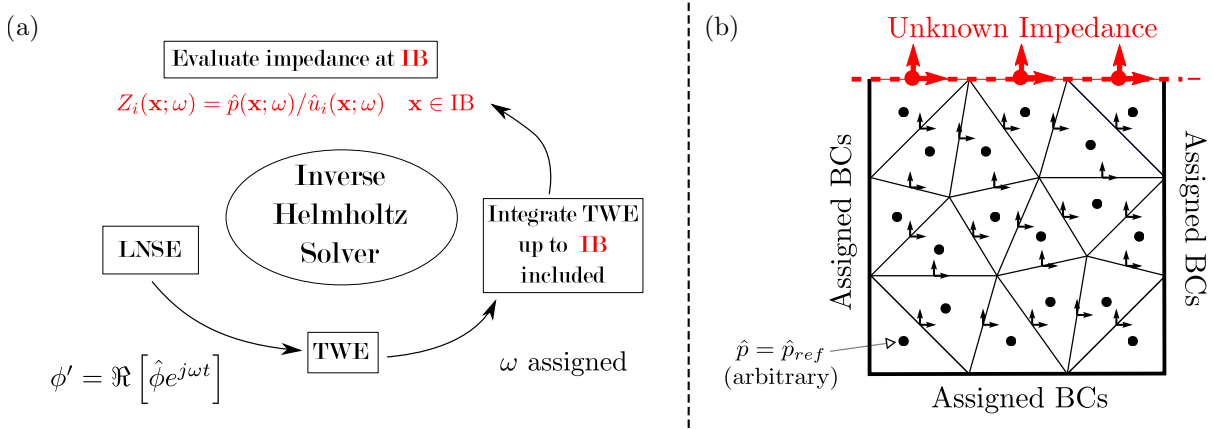

Figure 2: Conceptual illustration (a), and simplified two dimensional computational setup (b), of the inverse Helmholtz solver in two dimensions.

### III.D. Algebraic Closure of the inverse Helmholtz Problem

Closure of the iHS can be achieved via assignment of the surface gradient of pressure phase over the IBs. This corresponds to assigning the shape of the acoustic waveform crossing the IB and requiring only $n_{\text{IB}} - 1$ conditions to be specified. The remaining condition is the assignment of a nonzero value of the complex pressure amplitude, hereafter $\hat{p}_{ref}$ (figure 2b), at an arbitrary internal cell, which does not affect the resulting impedance at the IB due to the linearity of the TWEs.

Closure is achieved by assigning the quantity $\Psi_m$ defined as

$$\Psi_m = \frac{\hat{p}_{\text{IB},m+1}}{\hat{p}_{\text{IB},m}}, \quad \text{for} \quad m = \{1, \ldots, n_{\text{IB}} - 1\} \quad (10)$$

yielding the following closure conditions in matrix form,

$$\begin{pmatrix} \Psi_1 & -1 & 0 & \cdots & 0 \\ 0 & \Psi_2 & -1 & \cdots & 0 \\ \vdots & \ddots & \ddots & \ddots & \vdots \\ \vdots & \ddots & \ddots & \Psi_{n_{\text{IB}}-1} & -1 \\ 0 & 0 & 0 & 0 & 0 \end{pmatrix} \begin{pmatrix} \hat{p}_{\text{IB},1} \\ \hat{p}_{\text{IB},2} \\ \vdots \\ \vdots \\ \hat{p}_{\text{IB},n_{\text{IB}}} \end{pmatrix} = \begin{pmatrix} 0 \\ 0 \\ \vdots \\ \vdots \\ 0 \end{pmatrix}$$

or,

$$\underline{\underline{\mathcal{P}}}_{\text{IB}} \, \underline{\hat{p}}_{\text{IB}} = \underline{0} \tag{11}$$

Assigning an arbitrary reference pressure, $\hat{p}_{\text{ref}}$ to an arbitrarily chosen cell, following system is obtained,

$$\begin{pmatrix} 0 & \cdots & \cdots & \cdots & 0 \\ \vdots & \ddots & \ddots & \ddots & \vdots \\ 0 & \cdots & 1 & \cdots & 0 \end{pmatrix} \begin{pmatrix} \hat{p}_1 \\ \hat{p}_2 \\ \vdots \\ \hat{p}_{n_{\text{cv}}} \end{pmatrix} = \begin{pmatrix} 0 \\ \vdots \\ \hat{p}_{\text{ref}} \end{pmatrix}$$

or,

$$\underline{\underline{\mathcal{P}}}_{cv} \underline{\hat{p}} = \underline{b}_{\text{IB}}. \tag{12}$$

Combining the two closure conditions developed in equations (11) and (12), we get,

$$\underline{\underline{\mathcal{P}}}_{cv} \, \underline{\hat{p}} + \underline{\underline{\mathcal{P}}}_{\text{IB}} \, \underline{\hat{p}}_{\text{IB}} = \underline{b}_{\text{IB}}. \tag{13}$$

Replacing the $\underline{\underline{\phi}}_{\text{IB}}$ block in (9) by (13), we arrive at the underying system of equations for the iHS. Upon solving this system, the impedance of each IB face can be evaluated along the direction of the $i$−th velocity component as,

$$Z_{i,\text{IB}} = \frac{\hat{p}_{\text{IB}}}{\hat{u}_{i,\text{IB}}}. \tag{14}$$

### III.D.1. Planar Incident Wave Approximation

Unless otherwise specified the results shown in this manuscript are generated assuming planar incident waves in a quiescent medium i.e. $p_{\text{wave}} = A \exp\left(-ik\left(x\cos(\theta_n) + y\sin(\theta_n)\right)\right)$ such that the the complex pressure ratio (10) for the closure conditions (11) takes the form,

$$\Psi_m = \exp\Big(-i\{k\cos(\theta_n)(x_{\text{IB, m+1}} - x_{\text{IB, m}}) + k\sin(\theta_n)(y_{\text{IB, m+1}} - y_{\text{IB, m}})\}\Big), \tag{15}$$

where $k = \omega/a_0$ is the wave number, and $\theta_n$ the counter-clockwise angle made by the impedance boundary with the normal to the wave. For normal incidence, i.e. $\theta_n = \pi/2$, (15) further reduces to,

$$\frac{\hat{p}_{\text{IB, m+1}}}{\hat{p}_{\text{IB, m}}} = \Psi_m = 1. \tag{16}$$

## IV. Impedance of a thermoviscous straight duct

In this section the iHS methodology is applied to thermoviscous rectangular and circular ducts. Comparison with Rott's quasi-one-dimensional thermoacoustic theory[13,14] for normal wave incidence is first presented in §IV.B. The iHS is then used to compute normal and tangential impedance of a rectangular straight duct for the case of oblique wave incidence in §IV.C.

## IV.A. Duct Geometry and Base State Conditions

Figure 3 shows a representation of a two dimensional computational set up utilized for validation of the iHS methodology against Rott's linear theory for thermoviscous waves in rectangular ducts. For axisymmetric ducts, only the top half of the computational mesh was utilized with a symmetric (slip/adiabatic) wall at $y = 0$. Impedance was computed at the IB (figure 3) for homogeneous thermodynamic base-state,

$$p_0 = 101325 \text{ Pa}, \qquad \rho_0 = 1.2 \text{ kg/m}^3, \qquad T_0 = 293.15 \text{ K}, \qquad \gamma = 1.4, \text{ and} \qquad a_0 = \sqrt{\gamma R T_0} \text{ m/s}.$$

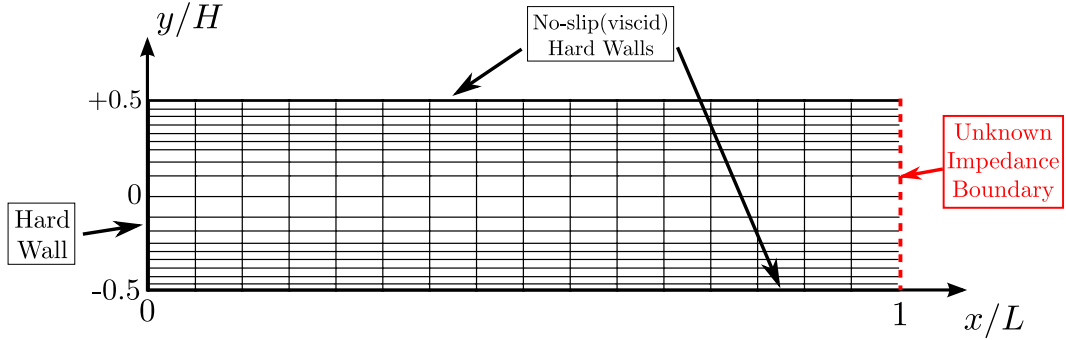

Figure 3: Computational set up (with an illustration of the mesh) to analyze a simple duct using the inverse Helmholtz Solver (iHS).

As previously stated in § III.D.1, present validation is restricted to planar incident waves. In general, planar wave approximation corresponds to far field propagation from a point source or arbitrarily incident waves with frequencies smaller than the cut off frequency $f_c$ of the duct, $f_c = a_0/2H$. Defining an axial Helmholtz number $He_L$ for the duct,

$$He_L = \frac{fL}{a_0} = \frac{fH(AR)}{a_0}, \tag{17}$$

admissible values of $He_L$ ensuring planarity of incident waves are given by, $He_L \leq AR/2$, where, $AR = L/H$ is the duct aspect ratio. For the constant aspect ratio of $AR = 4$ that is adopted to generate results in subsequent sections, this condition reduces to $He_L \leq 2$, indicating the validity of discussed results for non-planar waves with sufficient lower frequency as well.

## IV.B. Normally Incident Planar Wave

Closure conditions used in this subsection correspond to normally incident planar waves given by (16). A refined mesh with grid points concentrated near the boundaries was considered (figure 3) to obtain boundary layer resolved results. At lower frequencies (figure 4a), the iHS results show a deviation from the semi-analytical solutions. In Rott's quasi-1D formulation, wall-normal pressure-gradients and velocity ($\partial \hat{p}/\partial y$ and $\hat{v}$) are neglected compared to axial gradients and velocity ($\partial \hat{p}/\partial x$ and $\hat{u}$) based on a high aspect ratio assumption. However, sufficiently near the solid boundary, both $\hat{u}$ and $\hat{v}$ are of comparable magnitude hence resulting in curved streamlines which are captured in the 2D formulation of iHS indicating more accurate predictions at lower frequencies. As the frequency increases (figure 4f), local Stoke's layer thickness decreases, as a result of which the region confining the curved streamlines shrinks thus further approaching the quasi-1D approximation. Consequently, the results from the iHS and Rott's theory approach to identical values as the frequency increases.

To evaluate the impedance of axisymmetric ducts, the system of equations (9) was recast in cylindrical coordinates. Figure 5 shows the comparisons of dimensionless admittance, $Y_{*,\text{IB}}$, obtained from the iHS, with semi-analytical solutions obtained from Rott's wave equations for an axisymmetric duct.

## IV.C. Obliquely Incident Planar Wave

Closure conditions for a planar wave incident at an arbitrary angle $\theta_n$, towards the IB and are given by equation (15) where $\theta_n$ is the counter-clockwise angle between the normal to the planar wave and the impedance boundary, i.e., $\theta_n = 0$ is a grazing flow and $\theta_n = \pi/2$ is normal incidence.

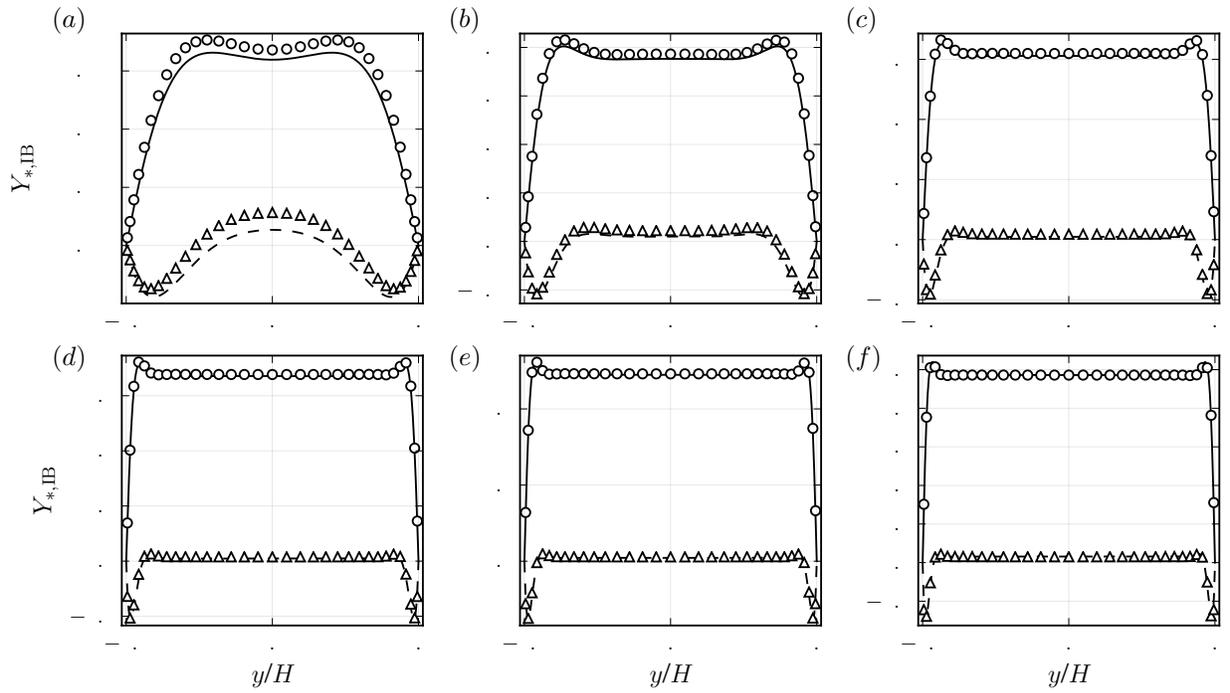

Figure 4: Comparisons of specific admittance obtained from the iHS ($\triangle$ for real, $\circ$ for imaginary) and Rott's wave equations ($--$ for real, $—$ for imaginary) for a two-dimensional rectangular viscous duct at dimensionless frequencies of $He_L(\times 10^3) =$ (a) 2.33, (b) 11.63, (c) 46.54, (d) 93.07, (e) 139.61, and (f) 186.14.

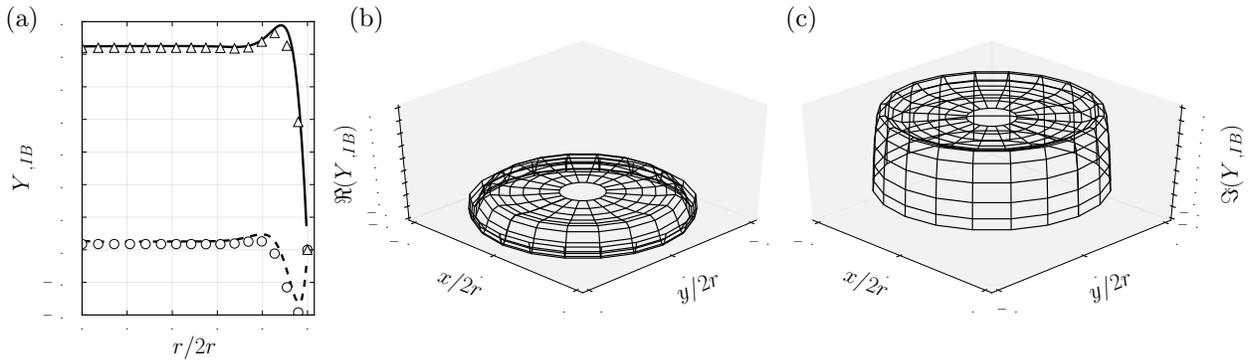

Figure 5: (a) Comparisons of specific admittance obtained from the iHS ($\triangle$ for real, $\circ$ for imaginary) and Rott's wave equations ($\cdots$ for real, $—$ for imaginary) for an axisymmetric duct at dimensionless frequency of $He_L(\times 10^3) = 46.54$. Figures (b) and (c) portray the real and imaginary parts of specific admittance obtained from the iHS in three dimensions.

Due to the symmetry of the duct considered, only cases with $\theta_n \in \left(0, \frac{\pi}{2}\right)$ are presented in figure 6 for $He_L = 46.54 \times 10^{-3}$. The pressure oscillations at the no-slip walls are in opposite phase indicating standing waves oscillating at the imposed frequency and wave number of $k\cos\theta_n$. Large values of $\Re(Y_{*,IB})$ indicate high momentum diffusion near the walls. Moreover, since the frequency is smaller than the cut-off frequency, the wall normal (standing) modes are evanescent and decay exponentially along the duct. The tangential admittance (figure 6b) is evaluated along the direction of positive $x$ axis. In general, the peak magnitudes of real and imaginary components of tangential admittance are smaller than their normal admittance counterparts. Further, the peak magnitude of the real component of admittance decreases with increasing angle of incidence. At normal incidence, $\theta_n = \pi/2$, the real component of tangential admittance vanishes completely.

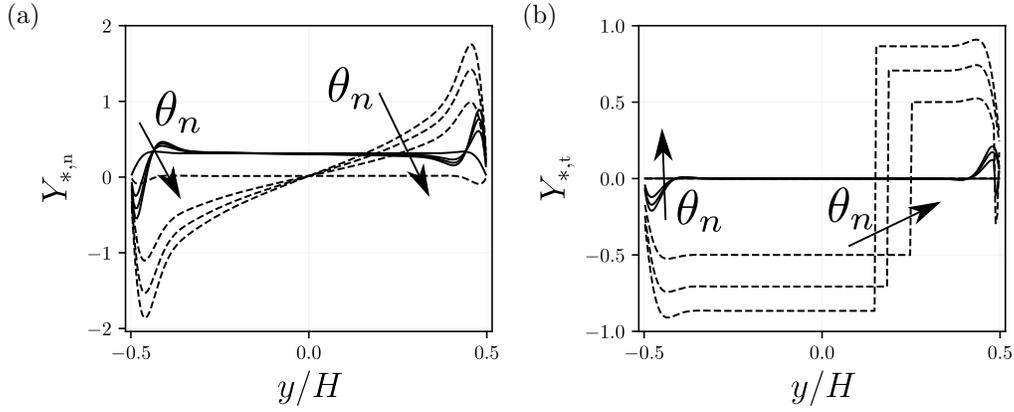

Figure 6: Spatial profiles of real (- - -) and imaginary (—) components of normal (a) and tangential (b) specific acoustic admittance for viscous planar wave propagation at angles of incidence, $\theta_n = \frac{\pi}{6}, \frac{\pi}{4}, \frac{\pi}{3},$ and $\frac{\pi}{2}$.

## V. Impedance of a toy porous cavity

In this section a geometrically complex toy cavity is investigated using the iHS. In §V.A, impedance at the open end of the toy cavity obtained from the iHS is compared with that obtained from fully compressible Navier-Stokes simulation of a monochromatic wave incident normally on the cavity. Finally, the application of iHS methodology in modeling acoustic cavities as lumped impedance boundaries[17] is discussed (§V.B). To this end, a narrowband pulse reflection is studied, first with resolved cavity and then modeled as lumped impedance boundary with frequency domain impedance of the cavity transformed into Time Domain Impedance Boundary Condition (TDIBC) following the procedure in Scalo et. al 2015.[16] Geometry and a representational unstructured mesh of the analyzed 2D toy cavity along with a computational set up for harmonic wave excitation are shown in figure 7. Momentum and thermal perturbations vanish near the circular obstacles (no-slip isothermal boundaries) and oscillating boundary layers develop thus mandating very small grid spacing for boundary layer resolved calculations of narrowband acoustic waves interacting with the cavity (§V.B). Usually, for instance in calculations of hypersonic boundary layer transition control, length scales of such cavities are very small compared to the entire domain as a result of which boundary layer resolved calculations inside the cavity pose high computational costs. However, impedance at the open end of the cavity (unknown impedance boundary, IB), if known, can be utilized for modeling the cavity as TDIBC in practical simulations thus eliminating the need for boundary resolved calculations inside the cavity to capture the reflection/absorption of narrowband acoustic waves.

Test cases in the following subsections assume planar wave propagation and normal incidence in a viscous medium at atmospheric base pressure and temperature $T_0 = 300$ K for air, modeled as an ideal gas. Dynamic viscosity is evaluated using Sutherland's law using $\mu_\mathrm{ref} = 1.827 \times 10^{-3}$ kg m$^{-1}$s$^{-1}$ to reduce computational costs associated to Stokes layer resolution.

### V.A. Harmonic excitation

Figure 7 shows the computational set up used for extracting the impedance at the open end of the cavity (vertical red dashed line, IB) from steady state response of a harmonic (acoustic) excitation of the cavity. A symmetric isothermal boundary condition with harmonic variation in temperature was imposed at the left end while periodic boundary conditions were imposed on the lateral walls (dashed lines) of the waveguide to avoid viscous attenuation. At steady state, the acoustic pressure and velocity perturbations in the Fourier space $(\hat{p}, \hat{u})$ were extracted to evaluate the acoustic impedance. The geometry of the cavity results in non-planar pressure field at the IB causing small deviation between the iHS and fully compressible Navier-Stokes results (as shown by pressure field contours in figure 7.

### V.B. Narrowband pulse excitation

Utilizing the iHS results, a narrowband pulse reflection was studied with the cavity modeled as equivalent time domain impedance boundary condition (TDIBC) and compared with the cavity resolved calculations.

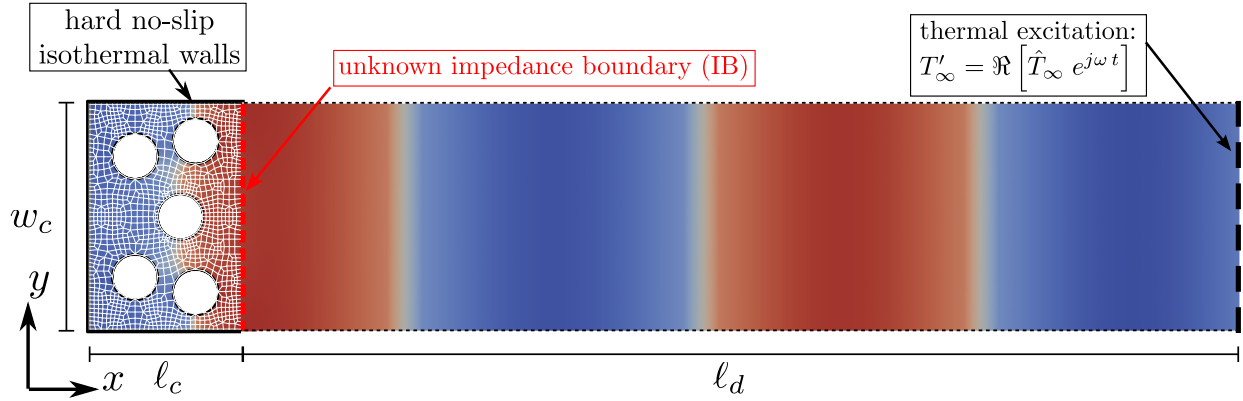

Figure 7: Pressure contours at steady state in the validation setup with monochromatic harmonic variation of temperature imposed at the left boundary (symmetric isothermal) at amplitude $|\hat{T}_\infty| = 0.1$ K.

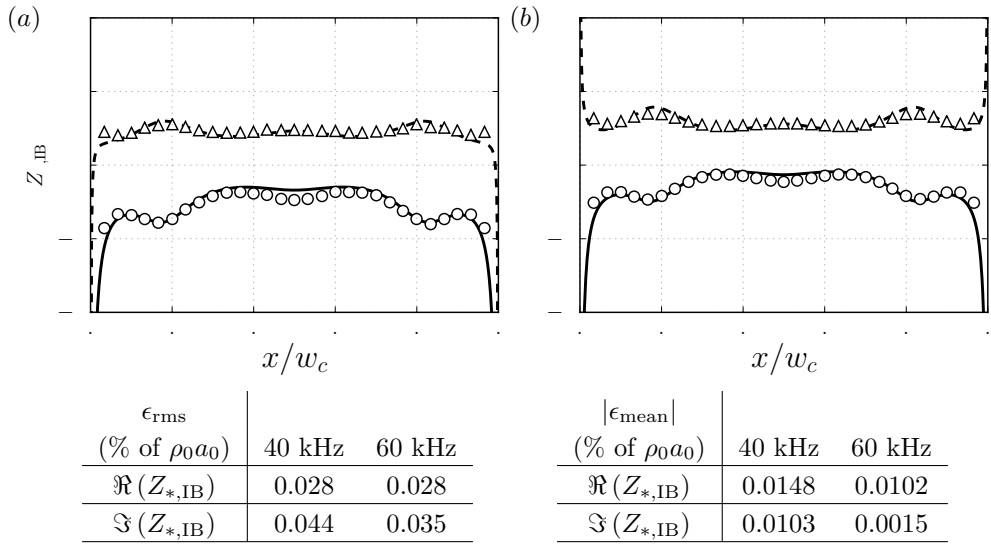

| $\epsilon_{\mathrm{rms}}$ (% of $\rho_0 a_0$) | 40 kHz | 60 kHz |
|---|---|---|
| $\Re(Z_{*,\mathrm{IB}})$ | 0.028 | 0.028 |
| $\Im(Z_{*,\mathrm{IB}})$ | 0.044 | 0.035 |

| $|\epsilon_{\mathrm{mean}}|$ (% of $\rho_0 a_0$) | 40 kHz | 60 kHz |
|---|---|---|
| $\Re(Z_{*,\mathrm{IB}})$ | 0.0148 | 0.0102 |
| $\Im(Z_{*,\mathrm{IB}})$ | 0.0103 | 0.0015 |

Figure 8: Dimensionless acoustic impedance obtained from the iHS (lines) and extracted from fully compressible unstructured harmonic excitation simulations (symbols) for frequencies, (a) 40 and (b) 60 kHz. Real part: ($--,\triangle$); Imaginary part ($-,\circ$).

Initial conditions (figure 9) for the two cases are given by,

$$p' = A_0\, e^{\left(-\frac{5}{4}k^2(y-\bar{y})^2\right)}\, \sin(2\pi k y), \quad v' = -\frac{p'}{\rho_0 a_0}, \quad \rho' = p'/a_0^2, \text{ and } T' = \frac{1}{\rho_0 R_{\mathrm{gas}}}p' - \frac{T_0}{\rho_0}\rho', \quad (18)$$

with,

$$A_0 = 5 \text{ Pa and } \bar{y} = 0.03 \text{ m}.$$

For modeling the cavity as TDIBC, surface averaged specific acoustic impedance from the iHS was fit to the three parameter model,[15]

$$Z_* = R + \left(X_{+1}\omega - X_{-1}\omega^{-1}\right) \quad (19)$$

in the frequency range 20–100 kHz, capturing the modes contained in the initial signal $p'$ given by equation (18) (figure 10). The considered initial pulse is composed of frequencies with corresponding wavelengths significantly larger compared to the characteristic dimensions of the cavity. As shown in previous section, in the frequency range of $10-100$ kHz, the planar wave approximation holds with very small errors introduced in surface average impedance values. Consequently, the pulse reflection in the present case is also approximated as one dimensional.

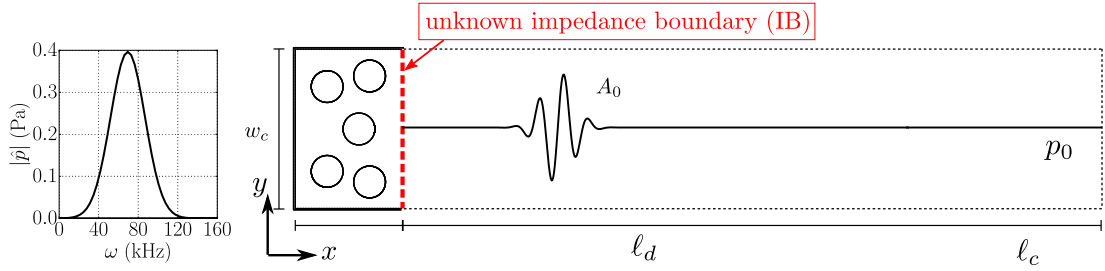

Figure 9: Initial set (not to-scale) up of the computational models used to compare the narrowband pulse reflection results from the cavity modeled as TDIBC (at the vertical dashed red line) with the cavity resolved Navier Stokes calculations. Spectral content of the narrowband acoustic pulse is shown on the right. Cavity dimensions: $l_d = 0.1$ m, $l_c = 0.001$ m, $w_c = 0.0015$ m.

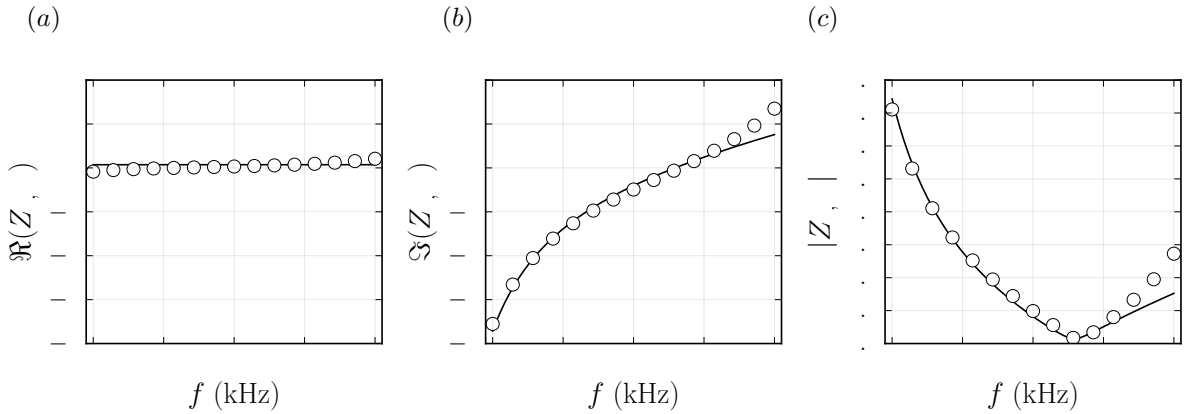

Figure 10: Comparisons between the (a) real part, (b) imaginary part, and (c) magnitude of the frequency-domain impedance obtained from the iHS (○) and approximations using the three parameter model (—).

Figure 10 compares impedance spectrum obtained from iHS to the single oscillator fit given by equation (19). An obvious limitation of the three parameter model is its inability to capture variations in $\Re(Z)$. An average is therefore applied over the entire range (figure 10a). The fit for the imaginary component of impedance is more robust, especially in the vicinity of $\Im(Z) \to 0$ (figure 10b). Overall, the fit retains necessary information regarding the frequency of peak acoustic absorption ($\sim 73$ kHz) and the approximate resistance ($R_n$) in the vicinity of this frequency (figure 10c). The calculated parameters, $R$, $X_{+1}$, and $X_{-1}$ were then used to model a time domain impedance boundary condition following the procedure in Scalo et. al.[16]

Figure 11a compares the spatial profiles of the reflected waves for the two cases at a fixed physical time. The dashed lines in the background correspond to the initial wave packet. The comparison of reflected pulse for the pore-resolved case with the initial wave packet suggests damping of high frequencies in the signal by the cavity and unattenuated reflection of the low frequency modes. The TDIBC case is able to effectively replicate this behavior, albeit with a root-mean-squared error of 5% when normalized by $A_0$ in the vicinity of the pulse. This error can be attributed to surface averaging of two-dimensional impedance and to the averaging of resistance, $R_n$ over the entire range of frequencies when fitting to the three parameter model (19). Considering these limitations, there is a good agreement between the results from the two cases.

Figure 11b compares the surface averaged instantaneous acoustic power flowing through the vertical (red) dashed lines in figure 9, which corresponds to the mouth of the cavity in the pore-resolved case, and the impedance wall in the TDIBC case. It can be seen that the area under the latter half of the instantaneous power curve is considerably smaller than the former half, implying that acoustic energy is absorbed by (or dissipated in) the cavity. The results from the two cases again show good agreement and the root-mean-squared error in the TDIBC case is found to be approximately 4.3% of the peak acoustic flowing through the mouth of the cavity at any given time through the reflection.

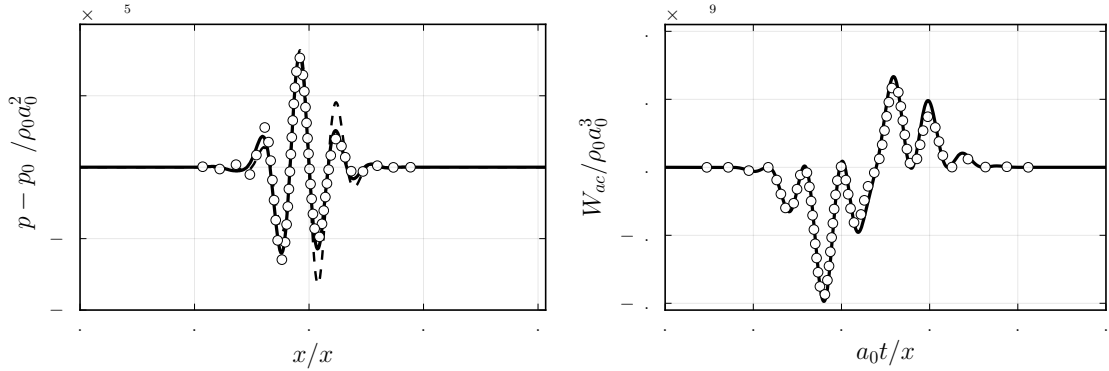

Figure 11: Comparison of TDIBC case (○) with full pore resolved case (—) via (a) spatial pressure profile of the reflected wave, and (b) instantaneous power flowing through a unit surface area at the impedance boundary (TDIBC case) and mouth of the cavity (pore resolved case). The dashed lines in the pressure plot correspond to the inital condition. Test case ran with parameters: $A_0 = 5$ Pa, $k = 200$ m$^{-1}$, $\bar{y} = 0.03$m.

## VI. Impedance of a thermoacoustically unstable cavity

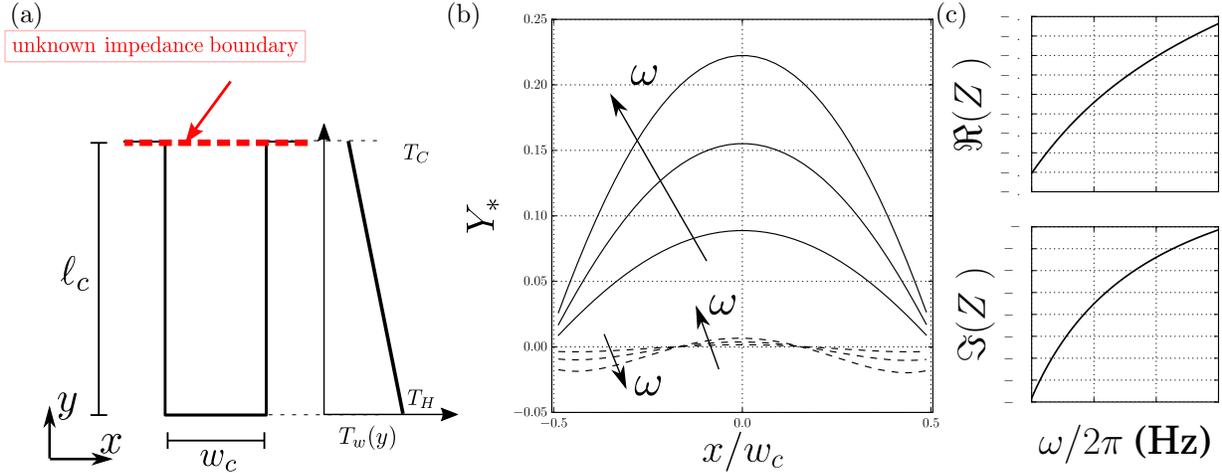

Figure 12: (a) Computational setup of the thermoacoustic cavity, (b) real (− −) and imaginary (—) components of complex acoustic admittance of the cavity for frequencies $\omega/2\pi = 1000$, 1750, and 2500 Hz, and (c) spectrums of real (top) and imaginary (bottom) components of complex acoustic impedance of the cavity between frequencies of $\omega/2\pi = 1000$ and 2500 Hz. Test cases were run with parameters $\ell_c$=4mm, $w_c$=0.1mm, $T_H$=800K, and $T_C$=300K.

Figure 12a shows the computational set up utilized in the analysis of a thermoacoustically unstable cavity with a linear temperature gradient applied at the isothermal walls and the bulk inside the cavity with the bottom wall at hot temperature ($T_H$) and the open end of the cavity at the cold temperature $T_C$. In order to obtain preliminary results, a constant viscosity at the mean temperature $(T_H + T_C)/2$ was computed using Sutherland's law. As shown in figure 12b Stokes' layer thickness, $\sqrt{2\nu/\omega}$ for the chosen conditions is comparable to the lateral dimension of the cavity. Furthermore, we see from figure 12c that the resistance, $\Re(Z_*)$, is negative, implying that the cavity produces energy, which has been shown from linear stability calculations involving the same cavity and temperature gradient as well.[18] Such an impedance prediction along with a multi-oscillator TDIBC is intended to be utilized in future work for efficient and less expensive (computationally) modeling of thermoacoustically unstable resonators in which boundary layer resolved calculations are mandatory for accurately capturing the energy density production inside the cavity (stack).

# VII. Conclusion

We have presented an inverse Helmholtz Solver (iHS) methodology to determine the linear acoustic impedance at one or multiple open boundaries (termed IB) of an arbitrarily shaped domain in response to an impinging wave of known frequency. In general, inverse Helmholtz problems are ill-posed and rank deficient, and need appropriate closure conditions. We show that assignment of the spatial distribution of the pressure phase over the IB, which could physically correspond to the shape of the incident wave, restores the rank of the system while serving as a closure condition (§III). Utilizing the proposed methodology, we calculated the impedance for open ended two-dimensional rectangular and circular ducts, with and without viscosity (§IV). These results were compared—and the methodology validated—against those obtained from Rott's quasi one-dimensional thermoviscous equations. In general, excellent matching was obtained for viscous ducts at high frequencies, owing to diminishing edge effects due to non-zero wall-normal gradients near the hard end of the duct, which are neglected in the derivation of Rott's wave equations. Following this, we analyzed an idealized geometrically-complex toy cavity using the iHS (§V). We first compared the specific acoustic impedance obtained from the iHS with that extracted from Navier Stokes simulations of harmonic excitations of the cavity. We then modeled the cavity as a lumped impedance boundary by transforming surface averaged impedance data from the iHS into a time domain impedance boundary condition (TDIBC), which was then applied to a Navier Stokes simulation of a narrowband pulse reflecting off of an impedance wall, and compared with equivalent Navier Stokes simulations wherein the full cavity is resolved. In spite of various approximations employed in the formulation of the TDIBC, the results from the two solvers are found to be in good agreement, highlighting the effectiveness in recovering impedance data using the iHS. Finally we analyzed a thermoacoustically unstable cavity yielding a preliminary evaluation of the broadband impedance spectrum of the cavity. Negative values of resistance at the impedance boundary imply thermoacoustic production of energy density, as has been shown from corresponding linear stability calculations as well. Such results can, in the future, be used to model the porous stack of a thermoacoustic engine by its equivalent mutli-oscillator time domain impedance boundary condition to reduce computational costs associated with stability studies of such devices. Other practical applications of the iHS include modeling of porous carbon-carbon coatings[19] that contain microporous cavities and are known to damp second-mode instabilities in hypersonic flows,[20,21] and modeling of high pressure distributor or choked nozzles adjacent to combustors that do not contain any reaction-zones but are pivotal to the stability of the latter.[22,23] Without significant modifications, the iHS can serve as a straightforward, standalone tool to evaluate the broadband impedance of such micro-scale cavities and choked nozzles, which can then be modeled as lumped impedance boundaries that retain necessary information, and implemented as TDIBC in solvers that only resolve the simplified (and reduced) domain. The advantage of such an approach would be seen in the form of significant reduction in computational costs at every iteration of the full time-domain simulation. Future work involves implementation of the iHS methodology in 3D, investigation of cases with mean flow and gradients, and application of the iHS to problems in hypersonics, thermoacoustics, and combustion.

# Acknowledgements

The authors acknowledge the support of the Rosen Center for Advanced Computing (RCAC) at Purdue and the Air Force Office of Scientific Research (AFOSR) grant FA9550-16-1-0209. The author Carlo Scalo is thankful for very fruitful discussions with Dr. Ivett Leyva (AFOSR).